\documentclass[fp,twocolumn]{jpsj3}
\usepackage{txfonts}
\usepackage[dvipdfmx]{color}

\title{Hybridization-gap Formation and Superconductivity in the Pressure-induced Semimetallic Phase of the Excitonic Insulator Ta$_{2}$NiSe$_{5}$}

\author{Kazuyuki~Matsubayashi$^1$\thanks{k.matsubayashi@uec.ac.jp}, 
Hidekazu~Okamura$^2$,
Takashi~Mizokawa$^3$,
Naoyuki~Katayama$^4$,
Akitoshi~Nakano$^4$\thanks{Present address: Department of Physics, Nagoya University, Nagoya 464-8602, Japan},
Hiroshi~Sawa$^4$,
Tatsuya~Kaneko$^5$\thanks{Present address: Department of Physics, Columbia University, New York, New York 10027, USA},
Tatsuya~Toriyama$^5$,
Takehisa~Konishi$^6$,
Yukinori~Ohta$^5$,
Hiroto~Arima$^1$,
Rina~Yamanaka$^7$,
Akihiko~Hisada$^{2,7}$,
Taku~Okada$^7$,\\
Yuka~Ikemoto$^8$,
Taro~Moriwaki$^8$,
Koji~Munakata$^9$,
Akiko~Nakao$^9$,
Minoru~Nohara$^{10}$,\\
Yangfan~Lu$^{11}$\thanks{Present address: Materials Research Center for Element Strategy, Tokyo Institute of Technology, 4259 Nagatsuta, Midori-ku, Yokohama 226-8503, Japan},
Hidenori~Takagi$^{11,12}$
and 
Yoshiya~Uwatoko$^7$
}
\inst{$^1$Department of Engineering Science, The University of Electro-Communications, Chofu, Tokyo 182-8585, Japan \\ 
$^2$ Graduate School of Technology, Industrial and Social Sciences, Tokushima University, Tokushima 770-8506, Japan\\
$^3$ Department of Applied Physics, Waseda University, Shinjuku, Tokyo 169-8555, Japan \\
$^4$ Department of Applied Physics, Nagoya University, Nagoya 464-8603, Japan \\
$^5$ Department of Physics, Chiba University, Chiba 263-8522, Japan \\
$^6$ Department of Chemistry, Chiba University, Chiba 263-8522, Japan \\
$^7$ Institute for Solid State Physics, The University of Tokyo, Kashiwa, Chiba 277-8581, Japan \\
$^8$ Japan Synchrotron Radiation Research Institute, Sayo 649-5198, Japan \\
$^9$ Comprehensive Research Organization for Science and Society (CROSS), Tokai, Ibaraki 319-1106, Japan \\
$^{10}$ Department of Quantum Matter, Graduate School of Advanced Science and Engineering, Hiroshima University, Higashi-hiroshima 739-8530, Japan \\
$^{11}$ Department of Physics, Graduate School of Science, University of Tokyo, Bunkyo-ku, Tokyo 113-0033, Japan\\
$^{12}$ Max Plank Institute for Solid State Research, Heisenbergstrasse 1, 70569 Stuttgart, Germany}

\abst{	
	The excitonic insulator Ta$_{2}$NiSe$_{5}$ experiences a first-order structural transition under pressure from rippled to flat layer-structure at $P_{\rm s}$ $\sim$ 3~GPa, which drives the system from an almost zero-gap semiconductor to a semimetal. The pressure-induced semimetal, with lowering temperature, experiences a transition to another semimetal with a partial-gap of $\sim$0.1-0.2~eV, accompanied with a monoclinic distortion analogous to that occurs at the excitonic transition below $P_{\rm s}$. We argue that the partial-gap originates primarily from a symmetry-allowed hybridization of Ta-conduction and Ni-valence bands due to the lattice distortion, indicative of the importance of electron-lattice coupling. The transition is suppressed with increasing pressure to $P_{\rm c}$ $\sim$ 8~GPa. Superconductivity with a maximum $T_{\rm sc}$ $\sim$ 1.2~K emerges around $P_{\rm c}$, likely mediated by strongly electron-coupled soft phonons. The electron-lattice coupling is as important ingredient as the excitonic instability in Ta$_{2}$NiSe$_{5}$.
}

\begin{document}
\maketitle

\section{Introduction}
	Excitonic insulator was theoretically proposed as a ground state of narrow gap semiconductors and semimetals \cite{Mott,Halperin,Bronold}. A spontaneous condensation of excitons with lowering temperature is expected to occur at a transition temperature $T_{\rm c}$ when the exciton binding energy $E_{\rm b}$ is larger than the one electron band gap $E_{\rm g}$. The state below $T_{\rm c}$ is called an excitonic insulator (EI). EI is most stable at around $E_{\rm g}$ $\sim$ 0. With increasing $E_{\rm g}$ from zero, EI is suppressed and disappears around $E_{\rm g}$ = $E_{\rm b}$. In a semimetal with  $E_{\rm g}$ $<$ 0, because of the increased screening, EI is suppressed rapidly with increasing $|E_{\rm g}|$. A dome-shaped $T_{\rm c}$-$E_{\rm g}$ curve peaked at around $E_{\rm g}$ $\sim$ 0 is expected. 1$T$-TiSe$_2$ and Tm(Se,Te) have been discussed very often as a candidate for EI \cite{Wilson,Cercellier,Wachter,Akiba}. To date, however, experimental evidences for excitonic state in these compounds are far from complete. 

	A new generation of candidate EI, a layered chalcogenide Ta$_{2}$NiSe$_{5}$ \cite{Sunshine,Salvo}, was very recently identified\cite{Wakisaka,Lu}. The layer consists of an array of alternating Ni single chain and Ta double chains running along the $a$-axis. Band structure calculations show that the valence band and the conduction band primarily comprise hybridized Ni 3$d$/Se 4$p$ states and Ta 5$d$ states, respectively \cite{Kaneko,Seki}.  Since both the top of valence band and the bottom of conduction band are located at the $\Gamma$ point, Ta$_{2}$NiSe$_{5}$ is a direct and almost zero-gap semiconductor/semimetal, which gives rise to an ideal playground for excitonic physics \cite{Okazaki, Mor, Sugimoto, Domon, Fukutani}. 
	
	An almost zero-gap semiconductor-to-insulator transition indeed occurs in Ta$_{2}$NiSe$_{5}$ at $T_{\rm c}$ $\sim$ 328~K, accompanied with a second-order structural phase transition from high-temperature orthorhombic ($Cmcm$) to low-temperature monoclinic ($C2$/$c$)\cite{Salvo}. This transition has been discussed to be a $q$~=~0 excitonic transition.  Optical conductivity $\sigma(\omega)$ indicates an opening of excitation gap of $\sim$0.2-0.3~eV below $T_{\rm c}$ = 328~K \cite{Lu}. The energy scale of gap is comparable to $E_{\rm b}$ $\sim$ 0.25~eV of the sister compound Ta$_{2}$NiS$_{5}$, which is estimated from the excitonic structure below the band edge in the optical conductivity spectrum $\sigma(\omega)$.  Ta$_{2}$NiS$_{5}$ has a much larger $E_{\rm g}$ $\sim$ 0.5~eV than Ta$_{2}$NiSe$_{5}$ and shows no signature of excitonic phase transition\cite{Larkin}. The control of one-electron gap $E_{\rm g}$ of Ta$_{2}$NiSe$_{5}$ was attempted, using a solid-solution Ta$_{2}$Ni(Se$_{1-x}$S$_{x}$)$_{5}$ \cite{Lu}. With the increase of $E_{\rm g}$ by S-substitution, $T_{\rm c}$ decreases and goes to zero temperature at around $x = 0.5$, where $E_{\rm b}$ $\sim$ $E_{\rm g}$ is expected. With the application of moderate pressure up to 3.0~GPa, $T_{\rm c}$ shows a decrease consistent with what is expected for the semimetallic side ($E_{\rm g}$~$<$~0) from $E_{\rm g}$ $\sim$ 0 \cite{Lu,Arima_1}. Combining the two results for $E_{\rm g}$ $>$ 0 and $E_{\rm g}$ $<$ 0 reproduces the canonical phase diagram as a function of $E_{\rm g}$. Those early observations support that the transition at $T_{\rm c}$ = 328~K is primarily excitonic in origin, which makes Ta$_{2}$NiSe$_{5}$ the most promising candidate material among EI candidates.
	
	Recently, however, it becomes increasingly clear that the actual situation is more complicated due to the simultaneous occurrence of monoclinic lattice distortion. While it is forbidden by symmetry in the undistorted orthorhombic phase, the hybridization of the Ta-conduction and the Ni-valence bands is allowed in the monoclinic phase. The structural distortion in the monoclinic phase is antiferroelectric-like, comprising the opposite displacement of the Ta chain and the surrounding Se ladder along the $a$-(chain) axis within the TaSe$_2$ and the asymmetric distortion of two neighboring TaSe$_2$ with respect to the Ni chain between the two \cite{Nakano,Nakano2}. Band calculations indicate that this antiferroelectric-like distortion couples electrons appreciably and gives rise to a sizable hybridization gap of $\sim$0.1~eV \cite{Subedi,Yaresko}, which is not small as compared with the excitonic gap of $\sim$0.2-0.3~eV. According to the theoretical studies, the symmetry requires that a finite excitonic order parameter must be accompanied with the monoclinic distortion \cite{Watson,Mazza}. The many-body excitonic gap, if any, is always mixed up with a one-electron hybridization gap. So far accumulated evidence for excitonic insulator points a strong excitonic character of the zero-gap semimetal to insulator transition in Ta$_{2}$NiSe$_{5}$. Nevertheless, the putative excitonic state cannot be a “pure” excitonic insulator because of the inherent admixture of hybridization gap character. 

	As is clear from the canonical $T_{\rm c}$-$E_{\rm g}$ phase diagram of excitonic insulator, excitonic instability should be suppressed rapidly on going from almost zero-gap region ($E_{\rm g}$ $\sim$ 0) to semimetallic region ($E_{\rm g}$ $<$ 0) because of the enhanced screening of Coulomb attraction between electrons and holes. If we can increase the overlap of Ta-conduction and Ni-valence bands and bring the system deeply inside the semimetallic region, the excitonic instability should be substantially reduced. The hybridization gap hidden behind the excitonic gap then may be uncovered. The hybridization gap is a one-electron effect and should not be so sensitive to the band overlap if the monoclinic distortion remains intact with the increase of band overlap. We may observe a change of character of gap from excitonic dominant to hybridization gap dominant under high pressures on increasing the band overlap.

	In this study, we explored the transport and the optical properties of Ta$_{2}$NiSe$_{5}$ under high pressures up to $\sim$9~GPa and present the complete phase diagram of high-pressure semimetallic phase above $P_{\rm s}$ $\sim$ 3~GPa, where a phase transition with a partial-gap is observed at $T^{\rm *}$, accompanied with a monoclinic distortion. We argue that the low temperature semimetallic phase is nothing but the hybridization-gap dominant phase produced by a strong electron-coupling with the antiferroelectric-like lattice distortion in Ta$_{2}$NiSe$_{5}$. $T^{\rm *}$ decreases rapidly with pressure and reaches to zero temperature around $P_{\rm c}$ $\sim$ 8~GPa. We discovered bulk superconductivity with transition temperature $T_{\rm sc}$ up to 1.2~K only around $P_{\rm c}$, which further justifies the strong electron-lattice coupling.

\section{Methods}
	Single crystals of Ta$_{2}$NiSe$_{5}$ were grown by a chemical vapor transport method using I$_2$ as the transport agent. Electrical resistivity and Hall resistivity were measured by a standard four- and five-probe technique, respectively. The electrical current was applied along the $a$-axis and the magnetic field was applied perpendicular to the $ac$ plane. The AC magnetic susceptibility was measured at a fixed frequency of 317 Hz with a modulation field of 0.1 mT applied along the $a$-axis. For the resistivity and AC magnetic susceptibility measurements, pressure was generated by a cubic anvil cell, which can produce quasi-hydrostatic pressures because of its multiple-anvil geometry \cite{Cheng}. Hall resistivity measurements were performed using an opposite anvil cell \cite{Kitagawa} with an outer diameter of 22~mm, installed in a Quantum Design Physical Property Measurement System (PPMS). For these transport measurements, glycerol was used as the pressure-transmitting medium. The applied pressure was calibrated by measuring the pressure dependence of superconducting transition temperature of lead. Note that the pressure determined at low temperature changes during heating due to the solidification of liquid pressure medium and the different thermal expansion coefficients of cell components, which does not exceed $\sim$0.6~GPa. The optical conductivity spectra, $\sigma(\omega)$, under high pressure were derived from the reflectance spectra, $R(\omega)$, measured with a diamond anvil cell (DAC) and the infrared synchrotron radiation at the SPring-8 facility. The as-grown surface of a single crystal was attached to the culet surface of a diamond anvil, and KBr was used as the pressure-transmitting medium in the DAC. Technical details on high-pressure infrared studies with DAC have been reported elsewhere \cite{Okamura}. Kramers$-$Kronig$-$constrained variational fitting \cite{Kuzmenko} was used to derive $\sigma(\omega)$ from $R(\omega)$ \cite{info}. The density-functional-theory-based electronic structure calculations were performed using the code WIEN2K \cite{WIEN}, based on the full-potential linearized augmented-plane-wave method. We adopted the generalized gradient approximation for electron correlations using the exchange-correlation potential \cite{Perdew}. In particular, we applied the mBJ exchange potential method \cite{Tran,Koller1,Koller2} to better describe the electron correlations. We set the parameter value of the mBJ potential $c$ to 1.9, ensuring that the band structures agree with a nearly zero-gap semiconductor at $P$ $<$ $P_{\rm s}$ and a semimetal at $P$ $>$ $P_{\rm s}$. This value is slightly larger than the value $c$~=~1.5 used in the previous publication \cite{Sugimoto}. We used the crystal structures and atomic positions measured under varying pressures at room temperature and focused on the differences in the electronic states between the low-$P$ $Cmcm$ and high-$P$ $Pmnm$ phases. In the self-consistent calculations, we used 144 (120) points from the irreducible part of the Brillouin zone for the $Cmcm$ ($Pmnm$) phase. The muffin-tin radii (R$_{\rm MT}$) of 2.43 (2.42), 2.22 (2.18), and 2.11 (2.08) Bohr radii were used for Ta, Ni, and Se ions in the $Cmcm$ ($Pmnm$) phase and the plane-wave cutoff of $K_{\rm max}$ = 7.00/R$_{\rm MT}$ was assumed.

\section{Results}

\begin{figure}[t]
\begin{center}
\includegraphics[width=0.48\textwidth]{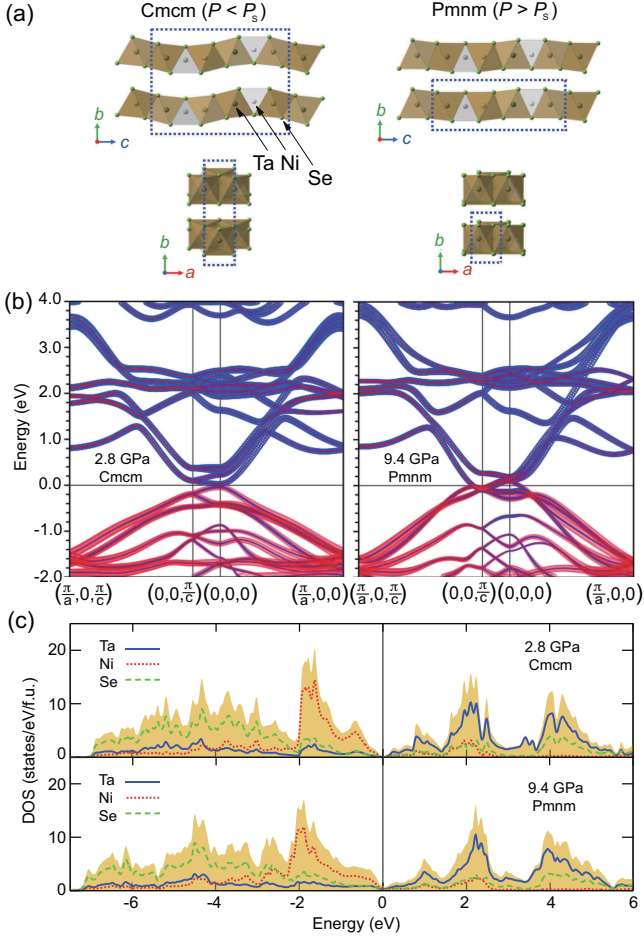}
\end{center}
\caption{(Color online) (a) Crystal structures of the low-$P$ orthorhombic phase ($Cmcm$) and the high-$P$ orthorhombic phase ($Pmnm$). The unit cells for low-$P$ and high-$P$ phases are indicated by broken lines. (b) Band dispersions and (c) total and partial densities of states calculated using the experimental crystal structures of the low-$P$ (at 2.8~GPa) and the high-$P$ (at 9.4~GPa) phases of Ta$_{2}$NiSe$_{5}$. The widths of the blue and red curves are proportional to the weights of the Ta and Ni contributions, respectively. The Fermi level is indicated by the straight line at $E$ = 0.
}
\label{Fig1}
\end{figure}

	Recent X-ray diffraction measurements under pressures revealed a first order structural phase transition from orthorhombic $Cmcm$ structure to a closely related orthorhombic $Pmnm$ structure at $P_{\rm s}$ $\sim$ 3.0~GPa \cite{Nakano}. Figure 1(a) displays the comparison of the low- and high- pressure structures. In the high-pressure $Pmnm$ phase, the unit cell contains only one layer of Ta$_{2}$NiSe$_{5}$ in contrast to the low-pressure $Cmcm$ phase in which the unit cell contains two layers. The layer structure with an alternating array of double Ta chains and Ni chain remains intact in the $Pmnm$ phases. While the layers are substantially rippled in the $Cmcm$ phase, the layer is flattened in the $Pmnm$ phase above $P_{\rm s}$, giving a small inter-layer distance which should be favored under a high pressure. The band dispersions and the electronic densities of states calculated for the high-pressure flat-layer phase are shown in comparison with those for the low-pressure rippled-layer (Fig.~1(b) and 1(c)). The overall band structure does not change appreciably as the rippled-layers in the low-$P$ phase are simply stretched in the high-$P$ phase. In the flat-layer phase, the overlap of the conduction band and the valence band is increased and the one electron gap $E_{\rm g}$ changes from almost zero to appreciable negative. It means that, because of the first order transition, we can explore deep inside the semimetallic ($E_{\rm g}$ $<$ 0) region above $P_{\rm s}$.

\begin{figure}[t]
\begin{center}
\includegraphics[width=0.48\textwidth]{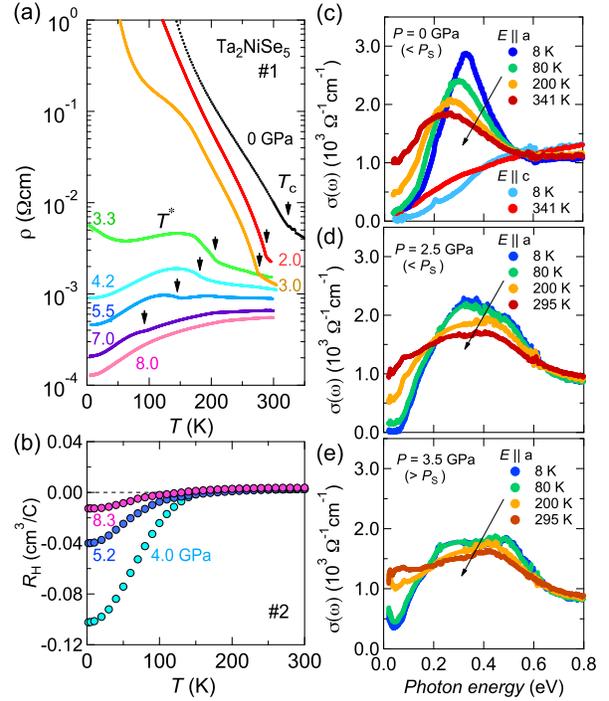}
\end{center}
\caption{(Color online) (a) Temperature dependence of the electrical resistivity $\rho(T)$ of Ta$_{2}$NiSe$_{5}$ ($\#$1) at selected pressures. A clear kink can be recognized in $\rho(T)$ at $T_{\rm c}$ and $T^{\rm *}$ below and above $P_{\rm s}$ respectively. $T_{\rm c}$ and $T^{\rm *} $ indicated by arrows are defined as the minimum of dln$\rho(T)$/d$T$.  (b) Temperature dependence of the Hall coefficient $R_{\rm H}$ of Ta$_{2}$NiSe$_{5}$ ($\#$2) above $P_{\rm s}$. (c) The real part of the optical conductivity $\sigma(\omega)$ with photon polarization along the chain direction ($E$ $||$ $a$-axis) and perpendicular to the chains ($E$ $||$ $c$-axis) at ambient pressure. The $a$-axis optical conductivity $\sigma(\omega)$ (d) at 2.5~GPa and (e) at 3.5~GPa.}
\label{Fig2}
\end{figure}

	With lowering temperature from above room temperature, $\rho(T)$ under low pressures up to $P_{\rm s}$ $\sim$ 3.0~GPa switches from a weakly semiconducting to a strongly insulating behaviors at a transition temperature $T_{\rm c}$ as shown in Fig. 2(a), reproducing the almost zero-gap semiconductor/semimetal-to-excitonic insulator transition. The kink-temperature, $T_{\rm c}$, is gradually suppressed with increasing pressure, indicative of the system moving slightly from $E_{\rm g}$ $\sim$ 0 to $E_{\rm g}$ $<$ 0 with pressure. With the pressure-induced first-order structural transition to the flat layer phase at $P_{\rm s}$ = 3.0~GPa, $\rho(T)$ in Fig.~2(a) discontinuously changes to a metallic behavior, which is fully consistent with previous reports \cite{Arima_1, Arima_2} and the result of band calculation, indicative of a semimetallic state with an increased band-overlap between the conduction and the valence bands. Note that $\rho(T)$ just above $P_{\rm s}$ weakly increases on cooling, which may be due to localization effect in a low-dimensional system. In this pressure-induced semimetal phase $P_{\rm s}$ $>$ 3.0~GPa, most importantly, we observe a signature of a phase transition as a kink in $\rho(T)$ at a temperature $T^{\rm *}$ lower than the excitonic transition temperatures $T_{\rm c}$ in the low-pressure rippled phase. A clear kink indicative of a phase transition is observed at $T^{\rm *}$ followed by a broad bump and then a fully metallic behavior, which is reminiscent of $T^{\rm *}$ for metals with a partial-gap formation associated with CDW and SDW transitions. As a central result of our work, other evidences described below indeed support the presence of a phase transition from a high temperature (HT-) semimetal to a low temperature (LT-) semimetal with a partial-gap at $T^{\rm *}$. With increasing pressure from $P_{\rm s}$, the phase transition temperature $T^{\rm *}$ shows a rapid decrease and the anomaly at $T^{\rm *}$ becomes substantially weaker than those at low pressures right above $P_{\rm s}$.  $T^{\rm *}$ appears to reach zero at around $P_{\rm c}$ $\sim$ 8~GPa, around and above which $\rho(T)$ becomes fully metallic without any anomaly.

	Previous x-ray study reported that the pressure induced flat-layer phase experiences a second-order structural transition from orthorhombic $Pmnm$ to monoclinic $P2/n$\cite{Nakano}, analogous to the case for the low-pressure rippled-layer phase. The monoclinic distortion comprises of antiferroelectric-like antisymmetric displacement of TaSe$_2$-unit as in the monoclinic distortion below the excitonic transition $T_{\rm c}$, which allows hybridization of Ta conduction and Ni valence bands. The reported structural transition temperature is 170~K for an applied pressure $P$ = 4.22~GPa \cite{Nakano}, which well agrees with $T^{\rm *}$ = 180~K for $P$ = 4.2~GPa in Fig.~2, and is suppressed with increasing pressure to $T$~=~0 around $P_{\rm c}$. Clearly, the $T^{\rm *}$ transition seen in $\rho(T)$ is accompanied with the monoclinic antiferroelectric-like lattice distortion.

	The Hall effect $R_{\rm H}$ and the optical conductivity $\sigma(\omega)$ measurements under a high pressure ($>$ $P_{\rm s}$) indicate that the LT-semimetal phase below $T^{\rm *}$ has a much lower carrier density than the HT-semimetallic phase and is accompanied with the formation of a partial gap at the Fermi level. The Hall effect $R_{\rm H}$ in the high temperature semimetal phase is positive and as small as $\sim$10$^{-3}$ cm$^{3}$/C likely due to a relatively large carrier concentration and a cancelation of electron and hole contributions, which is typical behavior for a semimetal at high temperatures. With lowering temperature to below $T^{\rm *}$, $R_{\rm H}$ gradually changes its sign to negative, indicative of electrons in the LT-semimetal phase having higher mobility than holes. The magnitude of $R_{\rm H}$ is enhanced appreciably to zero temperature limit. $R_{\rm H}$  $\sim$ 10$^{-1}$ cm$^{3}$/C in the low temperature limit at $P$ = 4.0~GPa corresponds to an electron density of $\sim$6$\times$10$^{19}$ cm$^{-3}$, which, considering the compensation of electron and hole contribution, should give an upper limit for the electron (hole) density in the LT-semimetal phase.
	
\begin{figure}[t]
\begin{center}
\includegraphics[width=0.48\textwidth]{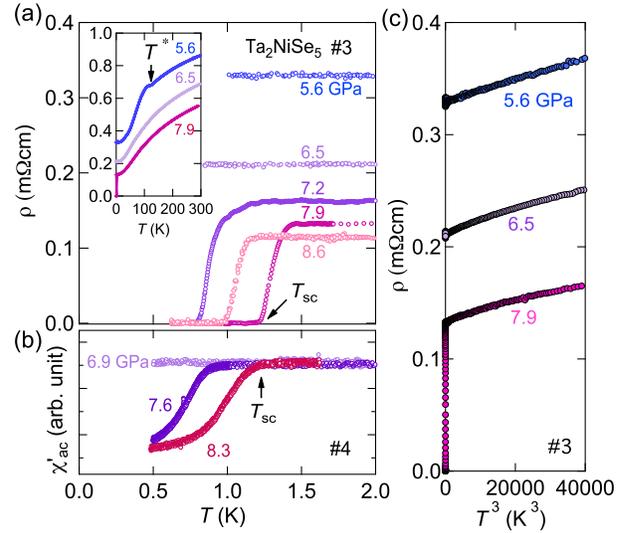}
\end{center}
\caption{(Color online) (a) Low-temperature $\rho(T)$ of Ta$_{2}$NiSe$_{5}$ ($\#3$) at pressures above $P_{\rm s}$. The inset shows temperature dependence of the electrical resistivity $\rho(T)$ over a wide temperature range. (b) AC magnetic susceptibility $\chi^{'}_{\rm ac}(T)$ of Ta$_{2}$NiSe$_{5}$ ($\#$4) under pressures. Here, the lead with almost the same volume and shape as the sample of Ta$_{2}$NiSe$_{5}$ was put inside the compensation coil to estimate the magnitude of the diamagnetic signal of Ta$_{2}$NiSe$_{5}$ by comparing to that of lead with opposite sign. The emergence of bulk superconductivity is confirmed by a large diamagnetic signal, corresponding to $\sim$40$\%$ of that for lead reference. (c) $\rho$ as a function of $T^{3}$ at selected pressures.
}
\label{Fig3}
\end{figure}
	
	The optical conductivity spectra $\sigma(\omega)$ under pressures point to that the reduction of carrier density in the LT-semimetal phase originates from the formation of partial gap at the Fermi level below $T^{\rm *}$. At ambient pressure, $\sigma(\omega)$ with photon polarization along the chain direction ($E$ $||$ $a$-axis) exhibits a characteristic peak centered at $\sim$0.3~eV assigned to the excitonic excitations, consistent with a previous report\cite{Lu,Seo}. With increasing pressure to $P$ = 2.5~GPa ($<$ $P_{\rm s}$), the excitonic peak broadens and the gap is reduced, indicative of the suppression of the excitonic state. This observation agrees well with the suppressed $T_{\rm c}$ in the resistivity measurements. $\sigma(\omega)$ with $E$ $||$ $a$ at $P$ = 3.5~GPa ($>$ $P_{\rm s}$) is quite different from that below $P_{\rm s}$. The pronounced peak around $\sim$0.3~eV is almost completely gone and a moderate temperature dependence is observed mainly below 0.1-0.2~eV.  While the low energy response is dominated by almost energy-independent $\sigma(\omega)$ above $T^{\rm *}$ = 200~K, a weak gap-like feature with an energy scale of 0.1-0.2~eV develops and, in addition, a small Drude-like component around $\omega$ = 0 is observed below $T^{\rm *}$.  The coexistence of the metallic Drude and the gap feature indicates that the bands around the Fermi level are gapped but only partly and that small Fermi surfaces remain below $T^{\rm *}$, which consistently accounts for the reduced carrier density inferred from the Hall effect measurements.

	We discovered bulk superconductivity at around the critical point $P_{\rm c}$. In another single crystal $\#$3 shown in inset of Fig.~3(a), $T^{\rm *}$ anomaly is observed up to 6.5~GPa but not clearly for 7.9~GPa. We extended our measurement below 4.2~K for crystal $\#$3. Fig.~3(a) shows the low temperature part of $\rho(T)$ data. At 7.2~GPa, a sudden drop of resistance to zero occurs around $T_{\rm sc}$ = 0.8~K, indicative of the emergence of superconductivity. $T_{\rm sc}$ reaches a maximum value of 1.2~K at 7.9~GPa and decreases with increasing pressure further to 8.6~GPa. This implies the presence of a dome-like superconducting phase around the critical point $P_{\rm c}$ = 8 GPa. The bulk nature of superconductivity can be evidenced by the AC susceptibility measurement on another crystal $\#$4 shown in Fig. 3(b), where the magnitude of diamagnetic signal at the lowest temperature is consistent with the perfect shielding. $\rho(T)$ in the $T$ = 0 limit shows $T^{3}$-dependence well below $P_{\rm s}$, however, the power of $T$-dependence becomes appreciably weaker than $T^{3}$ around $P_{\rm c}$ (see Fig.~3(c)), likely mirroring the presence of criticality.

\section{Discussion}
	The pressure-induced phase transitions observed in $\rho(T)$ are visually summarized as a $P$-$T$ phase diagram in Fig. 4. The first order transition from the rippled layer to the flat layer structures at $P_{\rm s}$ $\sim$ 3.0~GPa brings a contrast in the “pristine” electronic structure for orthorhombic structure without the monoclinic distortion between the two regions below and above $P_{\rm s}$, a narrow-gap semiconductor/semimetal below $P_{\rm s}$ and a semimetal above $P_{\rm s}$. With lowering temperatures, the narrow-gap semiconductor ($<$ $P_{\rm s}$) and the semimetal ($>$ $P_{\rm s}$) experience a phase transition to electronically distinct states, the excitonic insulator at $T_{\rm c}$ and to the LT-semimetal with a partial gap at $T^{\rm *}$ ($<$ $T_{\rm c}$) respectively. 
	
\begin{figure}[t]
\begin{center}
\includegraphics[width=0.45\textwidth]{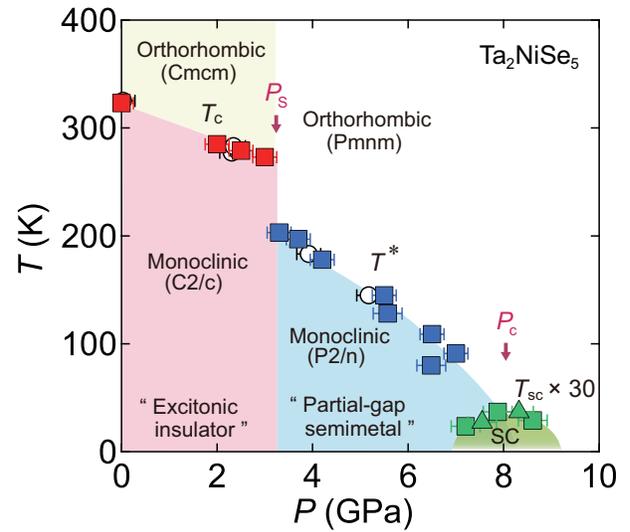}
\end{center}
\caption{(Color online) Temperature-pressure phase diagram of Ta$_{2}$NiSe$_{5}$. The phase transition temperatures $T_{\rm c}$ (below $P_{\rm s}$) and $T^{\rm *}$ (above $P_{\rm s}$) are determined from the $\rho(T)$ data (open circles are from Refs.~20 and 37.) Superconducting $T_{\rm sc}$ was determined as zero resistivity temperature in the $\rho(T)$ (squares) and the onset temperature of diamagnetic signal in the AC magnetic susceptibility  (triangles) data.}
\label{Fig4}
\end{figure}

	What is the main driver of the $T^{\rm *}$ transition to from the HT-semimetal to the LT-semimetal? We argue that the excitonic instability should be ruled out here. As there remain metallic carriers in the LT-semimetal below $T^{\rm *}$, the Coulomb interactions between electrons and holes should be screened out substantially in contrast to the case for the excitonic insulator phase below $P_{\rm s}$.  It is natural to ascribe the primary origin of the $T^{\rm *}$-transition under the absence of appreciable excitonic instability to the formation of hybridization gap associated with the monoclinic antiferroelectric-like lattice distortion, which is not a many-body effect but a one-electron effect. The recent band calculations indeed predict a hybridization gap of $\sim$0.1~eV in the monoclinic phase \cite{Subedi,Yaresko}, which is consistent with the energy scale of partial gap observed in the optical conductivity $\sigma(\omega)$ in the LT semimetal phase. We therefore conclude that the recently discussed physics of hybridization gap due to the monoclinic distortion in Ta$_{2}$NiSe$_{5}$ is unveiled in its high-pressure flat-layer phase by increasing the band overlap and suppressing the excitonic instability. The electron coupling to the antiferroelectric-like  lattice distortion is strong enough to give rise to a 0.1-0.2~eV gap, as pointed out theoretically. The appearance of superconductivity residing near the boundary of the suppression of the structural transition at $T^{\rm *}$ is in support of the strong electron lattice coupling. Around the critical point, the presence of low-lying soft phonon modes associated with the monoclinic transition which are coupled to strongly with electrons and very likely gives rise to a superconductivity. 
	
	It is tempting to infer that the energy scale of gap and the transition temperature $T_{\rm c}$ for the low-pressure excitonic insulator phase, $\sim$0.2-0.3~eV and $\sim$330 K, is larger but not by an order of magnitude than those of high-pressure semimetallic phase, 0.1-0.2~eV and 200~K, predominantly produced by the hybridization. This appears to suggest a substantial admixture of hybridization gap character even in the excitonic insulator phase below $P_{\rm s}$ $\sim$ 3~GPa. The electron-lattice coupling may be one of the key ingredients of the insulating gap formation below~$T_{\rm c}$.  
	
	The $\rho(T)$ of 1$T$-TiSe$_2$ under pressures \cite{Morosan,Kusmartseva} shows strikingly parallel behavior to that of the pressure induced semimetallic phase of Ta$_{2}$NiSe$_{5}$, including the kink anomaly at $T^{\rm *}$, the occurrence of superconductivity and the change of $\rho(T)$ from $T^{3}$- to a weaker temperature dependence at the $T^{\rm *}$ critical point. The pressure-induced semimetal phase in Ta$_{2}$NiSe$_{5}$ and the 1$T$-TiSe$_2$ should share the same ingredient of physics in their electronic structure and their electron-lattice coupling. Therefore, the main driving force of low temperature semimetal phase of 1$T$-TiSe$_2$ is highly likely a hybridization gap or CDW due to strong electron-lattice coupling. If the energy gap $E_{\rm g}$ could be tuned from negative to zero in 1$T$-TiSe$_2$ by a negative pressure, an insulator phase with dominant excitonic character as observed in Ta$_{2}$NiSe$_{5}$ would emerge.

\section{Summary}
	In summary, we have constructed the comprehensive phase diagram of the excitonic insulator Ta$_{2}$NiSe$_{5}$, deeply inside the semimetallic region. A semimetal-semimetal transition with lowering temperature is observed at $T^{\rm *}$, which is argued to be a hybridization gap formation due to the monoclinic lattice transition. The results evidence the presence of a strong electron-coupling with monoclinic distortion in Ta$_{2}$NiSe$_{5}$, which plays a vital role for the occurrence of superconductivity, common to 1$T$-TiSe$_2$, and likely even in the excitonic transition in the low pressure rippled phase.\\

\begin{acknowledgment}
	We thank K.~Okazaki, Y.~Fuseya, H. Matsuura, K.~Miyake, M.~Ogata, H.~Fukuyama, A.~Yaresko and A.~Rost for fruitful discussions, and K.~Kitagawa for experimental support and discussions. This work was partly supported by the Futaba Electronics Memorial Foundation, Japan Society for the Promotion of Science (JSPS) KAKENHI (No. 26400349, 15H05852, 15H03681, 17H01140, 17K05530, 18H04312, 18H01172, 19H00648 and 20H01849) and Alexander von Humboldt foundation. T.K. and T.T. acknowledge support from the JSPS Research Fellowship for Young Scientists. Experiments at SPring-8 were performed under the approval by JASRI (2013A1085, 2014B1749, 2014B1751, 2015B1698, 2015A1528, 2016A3782, 2016A1166). A part of this work were performed under the Shared Use Program of the QST Facilities (proposal No. 2016A-E17) supported by the QST Advanced Characterization Nanotechnology Platform of the Ministry of Education, Culture, Sports, Science and Technology (MEXT), Japan (proposal No. A-16-QS-0009). The use of the facilities of Coordinated Center for UEC Research Facilities and Cryogenic Center at University of Electro-Communications is appreciated.
\end{acknowledgment}

\bibliography{Reference}

\begin{thebibliography}{10}
\bibitem{info}
Due to strong absorption by diamond, the spectral range between 0.22 and 0.32
  eV could not be measured well. Therefore, this spectral range in R($\omega$)
  was interpolated smoothly before the Kramers-Kronig analysis.
\end{thebibliography}

\end{document}